\documentclass[prb,aps,showpacs,twocolumn,superscriptaddress]{revtex4-1}

\usepackage{graphicx}
\usepackage{ifthen}
\usepackage{ifpdf}
\usepackage{color}

\usepackage{latexsym}
\usepackage{amsthm}
\usepackage{amsmath}
\usepackage{amsfonts}
\usepackage{amssymb}
\usepackage{bm}
\newcommand{\half}{\mbox{\small $\frac{1}{2}$}}

\newcommand{\im}{\mbox{Im}}

\newcommand{\eexp}{\mbox{e}^}

\newcommand{\bea}{\begin{eqnarray}}
\newcommand{\eea}{\end{eqnarray}}
\newcommand{\expv}[1]{\left\langle {#1} \right\rangle}

\newcommand{\beq}[1]{\begin{eqnarray}\ifthenelse{#1=-1}{\nonumber}
{\ifthenelse{#1=0}{}{\label{e#1}}}}
\newcommand{\eeq}{\end{eqnarray}}
\newcommand{\be}[1]{\begin{eqnarray}}
\newcommand{\ee}{\end{eqnarray}}

\newcommand{\hide}[1]{}

\newcommand{\tr}{{\text Tr}}

\renewcommand{\vec}[1]{\mathbf{#1}}
\def\dd{\mathrm{d}}
\def\vrho{\boldsymbol{\rho}}
\def\vsigma{\boldsymbol{\sigma}}
\def\va{\vec{a}}
\def\hrho{\hat{\rho}}
\def\hA{\hat{A}}
\def\vb{\vec{b}}
\def\vB{\vec{B}}
\def\tvb{\tilde{\vb}}
\def\tvB{\tilde{\vB}}
\def\tvrho{\tilde{\vrho}}

\begin{document}

\title{Floquet topological phases coupled to environments  and the induced photocurrent}

\author{Szabolcs Vajna}
\affiliation{Department of Physics and BME-MTA Exotic  Quantum  Phases Research Group, Budapest University of Technology and Economics, 1521 Budapest, Hungary}
\affiliation{Department of Physics, Boston University, 590 Commonwealth Ave., Boston, MA 02215, USA}
\author{Baruch Horovitz}
\affiliation{Department of Physics, Ben-Gurion University of the Negev, Beer Sheva 84105 Israel}
\author{Bal\'azs D\'ora}
\affiliation{Department of Theoretical Physics and BME-MTA Exotic  Quantum  Phases Research Group, Budapest University of Technology and Economics, 1521 Budapest, Hungary}
\author{Gergely Zar\'and}
\affiliation{Department of Theoretical Physics and BME-MTA Exotic  Quantum  Phases Research Group, Budapest University of Technology and Economics, 1521 Budapest, Hungary}

 \pacs{ }
\begin{abstract}
We consider the fate of a helical edge state of a spin Hall insulator and its topological transition in presence of a circularly polarized light when coupled to various forms of 
environments. A Lindblad type equation is developed to determine the fermion occupation of the Floquet bands. We find by using analytical and numerical methods that non-secular terms, 
corresponding to 2-photon transitions, lead to a mixing of the band occupations, hence the light induced photocurrent is in general not perfectly quantized in the presence of finite coupling to the environment, 
although deviations are small in the adiabatic limit. 
Sharp crossovers are identified at frequencies $\Omega$ and $\half\Omega$ ($\Omega$ is the strength of light-matter coupling) 
with the former resembling to a phase transition.
\end{abstract}
\maketitle

\section{Introduction} \label{sec:intro}

Topological insulators (TI) are at the  focus of attention, representing peculiar states of matter with robust, topologically protected
conducting edge or surface states \cite{hasankane,qi}. Due to the strongly entangled spin and charge degrees of freedom, possible applications in spintronics or quantum computation
have been proposed. In particular, the two-dimensional TI, i.e. the
quantum spin-Hall (QSH) state has been predicted and experimentally observed for a number of systems, including graphene \cite{kanemele1}, HgTe/CdTe \cite{bernevig,konig} and
InAs/GaSb \cite{cxliu2008} quantum wells, lattice models \cite{weeks,guo,sun} and multicomponent ultracold fermions in optical lattices \cite{jiang,stanescu,goldman}.

While engineering topologically non-trivial band structures is far from being trivial, several methods have been proposed to induce TIs.
Among these,  time periodic driven quantum systems \cite{kitagawa} have been investigated by using Floquet theory\cite{sambe,shirley,dittrich}, the temporal analogue of Bloch states.
The resulting driven topological insulators are referred to as Floquet topological insulators.
 It has been proposed that novel topological edge states can be induced by irradiating electromagnetic waves on topologically
trivial material such as a non-inverted HgTe/CdTe quantum well~\cite{lindner} or simply graphene\cite{kitagawa} that has no topologically protected edge states in the absence of radiation.
Of further interest is the proposal by D\'ora et al. for a quantized photocurrent in a quantum spin Hall (QSH) and a topological phase transition
to a non-quantized photocurrent, when the frequency of the radiation field matches twice the energy of the Zeeman coupling, altering the topological properties \cite{dorafloquet}.
Besides the theoretical appeal of Floquet TIs, the Floquet shadow bands on the surface of a 3-dimensional TI Bi$_2$Se$_3$ have  been observed \cite{gedik} experimentally.
In addition, photonic waveguides have been used to simulate graphene interacting with circularly polarized light, and the existence of edge states
was revealed\cite{rechtsman}.

The steady state of Floquet topological insulators,  is described by the Floquet theory. 
Although the resulting Floquet spectrum often possesses a topology different  from that of their static parents,
the actual occupation of the various Floquet bands is, however, essential to  evaluate physical observables.
For example, a topologically non-trivial but only partially filled band cannot profit from topological protection.
The  occupation of the Floquet bands is, in principle, determined by the sources of relaxation, e.g. coupling to heat baths and phonons, 
momentum scattering from static disorder, or interparticle interaction. 
In their absence, one can borrow from the  Floquet literature\cite{dittrich} and assume  fermion occupations, 
which minimize the time averaged Hamiltonian, as was done, e.g.,  in Refs.~\onlinecite{dorafloquet,zhouwu}.

\begin{figure}[b!]
\centering
\includegraphics[width=8cm]{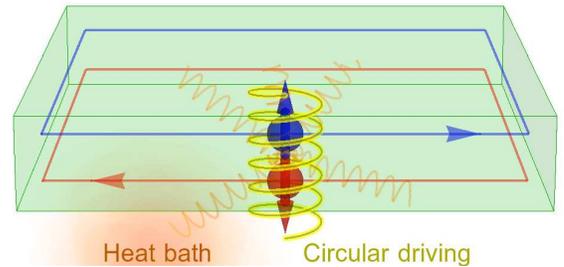}
\caption{The cartoon of the system, consisting of a QSH edge state with spin filtered conducting channels, interacting with circularly polarized electromagnetic field and coupled to an environment, is visualized.}
\label{fig:system}
\end{figure}

In the present work  we extend the model for a driven QSH system \cite{dorafloquet} to include various types of 
environments. 
In particular, we study a QSH insulator coupled to a bosonic heat bath, and irradiated by a circularly polarized light (see Eq.~\eqref{e15}). 
The electromagnetic field acts as a periodic driving as it couples the QSH edge states. The system is schematically sketched in Fig. \ref{fig:system}. 
For the sake of simplicity, we consider a model with the simplest possible form of a bosonic  dissipation, where dissipation does not 
couple states of different momenta, but drives  spin flip transitions.

\begin{figure}
\centering
\includegraphics[width=8.4cm]{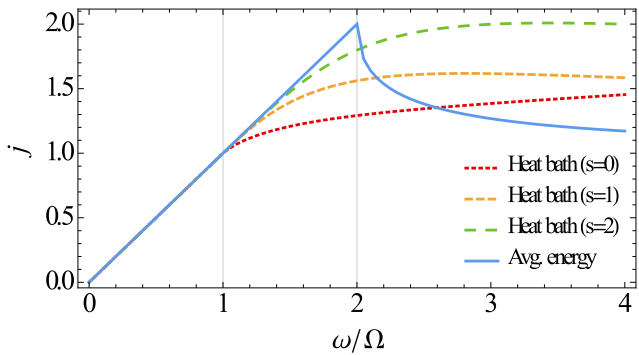}
\caption{Comparison of the edge current (in units of $e\Omega/2\pi)$) when the states are occupied based on their average energy \cite{dorafloquet}, and when they are coupled to a zero temperature bath.
The curves correspond to the secular approximation, which describes the infinitesimal system-bath coupling. The $s=0$ curve is understood as the limiting behavior as $s\rightarrow 0$.}
\label{fig:curravg}
\end{figure}

Following the lines of Ref.~\onlinecite{kamleitner}, we apply a generalized Lindblad type formulation (the Bloch-Redfield equations) to 
describe how the environment affects the dynamics of the edge states. In particular we keep non-secular terms, which are not captured in the Lindblad equation, but are found 
to affect the dynamics considerably. This requires, in general, a numerical solution, 
though near critical points we find that there is a single dominant non-secular term that allows a rotation into a time independent frame.
We find that the occupation of the bands deviates from the one found using the average energy assumption,\cite{dorafloquet,zhouwu} 
which leads to a \emph{weak violation} of \emph{current quantization} in the Floquet topological phase. Our main result concerning the induced photocurrent along the edge is summarized in
Fig. \ref{fig:curravg}.
 
Other Floquet systems have also been studied in the presence of disorder or dissipation\cite{dehghani,kundu,foatorres,farrell,XuPRA2014,IadecolaPRB2013,IadecolaPRB2015,IadecolaPRB2015b,DehghaniArXiv2015}.
Most of the former studies were related to driven graphene, showing that dissipation effects generally inhibit the naive generalization of the static results on topological band structures to the Floquet 
case, due to the non-thermal occupation of these bands. The effect of non-secular terms in graphene shined by circularly polarized light was studied numerically after a quantum quench of the driving field in  
Ref.~\cite{DehghaniArXiv2015}, our method 
of analytical treatment of the dominant resonances generalize to that case too, albeit we do not consider quenches here.

After introducing the model and the Floquet solution in the absence of heat bath in Sec II.,  we develop the Lindblad formulation in Sec. III. and  apply this formulation to the edge states of a driven QSH system in Sec. IV. 
We show numerical and approximate solutions to the fermion occupation, then discuss the stationary edge current in section V.

\section{Dissipative chiral edge states with a periodic drive}

We consider a one-dimensional chiral edge state of a QSH insulator in a circularly polarized radiation field, shown in Fig. \ref{fig:system} 
described by the non-interacting Hamiltonian $\tilde{\cal H}_{S} = \sum_p \psi^\dagger_p  \,\tilde{\cal H}_{S}(p)\,\psi_p$, with
\bea
\tilde{\cal H}_{S}(p)&=&\half p\sigma_z -\half\Omega(\sigma_+\eexp{-i\omega t}+h.c.).
\label{e15}
\eea
Here $ \psi^\dagger_{p,\sigma}$ creates a SQH edge excitation of momentum $p$ and spin $\sigma$, 
with $\half p$ is the energy  of the right moving spin up fermions, and $-\half p$ that of the left moving spin down
fermions (Fermi velocity is set to 1/2). The term with $\Omega$ comes from the Zeeman coupling between the 
magnetic component of the $\omega$ frequency electromagnetic field and the electron's spin,  and $\Omega$
is identified as the Rabi frequency.

For the sake of simplicity, we shall assume in the following that excitations of the environment have a very long wavelength 
compared to that of  edge excitations, and will also neglect the coupling it generates 
between different momenta. Under these conditions, we can restrict our considerations 
to a single momentum mode $p$, which  we then couple to the environment through
\begin{gather}
\tilde{\cal H}_{SE}=-\half b_x \sigma_x X -\half b_y \sigma_y Y-\half b_z\sigma_z Z\;.
\label{e15a}
\end{gather}
Here $X$, $Y$ and $Z$ denote Gaussian bosonic fields, coupled to the 
Pauli matrices,  and $b_{\mu}$ ($\mu\in\{x,y,z\}$) denote the corresponding couplings.
Their dynamics is encoded in 
the environment Hamiltonian, $ \tilde{\cal H}_E=\tilde{\cal H}_E(X,Y,Z)$, 
whose explicit form is not needed here as it only determines the spectral functions of the 
noise.   
We refer to this coupling scheme as the XYZ coupling. Below we consider also other forms of $\tilde{\cal H}_{SE}$, which are given by
identifying Y with X (the XXZ scheme), and both Y and Z with X  (referred to as XXX coupling). 

The actual form of the system bath coupling depends
on the physical realization, but as we will show, in the limit of weak coupling, they give similar results.
The environment is characterized by the bath spectral functions $J_{\mu=x,y,z}(\omega)=\alpha \omega_c^{1-s}\omega^s \eexp{-\omega/\omega_c}$, 
which determine the correlation functions $\gamma_{\mu}(\omega)=\frac{\eexp{\beta \omega}}{\eexp{\beta \omega}-1}J_{\mu}(\omega)$ at arbitrary 
temperature ${1}/{\beta}$. The dimensionless quantity $\alpha$ is the spectral strength and $\omega_c$ is a high frequency cutoff. An Ohmic 
bath corresponds to $s=1$, while $s\lessgtr 1$ describes the sub- and super-Ohmic baths, respectively.

We reemphasize that,  in our simplified model, 
each $p$ mode in Eq. \eqref{e15} is coupled to a different environmental variable, and 
similarly to Ref.~\onlinecite{dehghani}, the environment induced scattering between different 
momentum states is neglected, an assumption that simplifies the description 
of the resulting state considerably.

Let us start by reviewing the Floquet solution of the uncoupled topological insulator and its basic properties \cite{dorafloquet}.
The time-dependent Schr\"odinger equation,
$i\partial_t\Psi_p(t)=\tilde{\cal H}_S(p)\Psi_p(t),$
is solved using the Floquet ansatz\cite{sambe,dittrich} for the steady state solution, 
\begin{gather}
\Psi_p(t)=\exp[-i E_{\pm}(p)t]\Phi_{\pm}(p,t).
\label{floquetwavefunction}
\end{gather}
Here,  $E_{\pm}(p)$ denotes the Floquet quasienergy, and $\Phi_{\pm}(p,t)=\Phi_{\pm}(p,t+T)$ with $T=2\pi/\omega$,
\begin{gather}
E_{\pm}(p)=\frac{\omega\pm\Omega'}{2},
\label{fqe}\\
\Phi_{\pm}(p,t)=\frac {1}{\sqrt{2\Omega'}}
\left(
\begin{array}{c}
\sqrt{{\Omega' \mp \delta\omega}}\\
\pm \exp(i\omega t)\sqrt{{\Omega' \pm \delta\omega}}
\end{array}\right),
\label{wf}
\end{gather}
where  $\Omega'=\sqrt{\delta\omega^2+\Omega^2}$ is the renormalized Rabi frequency 
 and $\delta\omega=\omega-p$ denotes the detuning.
The Floquet quasienergies \eqref{fqe}
are only well defined modulo $\omega$ and cannot be used to determine the filling of Floquet states.
A phenomenological way of determining filled Floquet states \cite{zhouwu} relies on the  average energy,\cite{dittrich}  defined as 
\begin{gather}
\bar E_{\pm}(p)=\frac{1}{T}\int\limits_0^T{\rm d} t\, \Psi_p^+(t)H\Psi_p(t)=\frac{\pm}{2}\left[\Omega'-\frac{\omega\delta\omega}{\Omega'}\right],
\label{meanenergy}
\end{gather}
which is always single valued as opposed to the ladder of quasienergies in Eq.~\eqref{fqe}.
In the present work, we go beyond this phenomenological reasoning and determine the filling of the Floquet eigenstates from first principles.

\section{The non-secular Lindblad equation}

In this Section, we follow the lines of Refs. \onlinecite{shnirmanbook} and \onlinecite{kamleitner} 
to outline a theoretical framework to 
describe  the time dependent reduced density matrix by a generalized Lindblad-type equation,  
up to 2nd order in the coupling with an environment.

As a first step, one switches to the interaction picture with respect to the 
non-interacting Hamiltonians, ${\cal H}(t) = \tilde{{\cal H}}_S(t) + \tilde{{\cal H}}_{E}$,  
where the time evolution of the interacting system's density matrix $\rho(t)$ is 
governed simply  by the Hamiltonian ${\cal H}_{SE}(t)$ which we factorize as
\begin{equation} 
{\cal H}_{SE}(t)=\sum_\mu A_\mu(t)\otimes B_\mu(t).
\end{equation}
Here the operators $A_\mu(t)$ and $B_\mu(t)$ act on the system and the environment, 
and their time evolution is governed by $\tilde {\cal H}_S(t)$ and $\tilde {\cal H}_E$, respectively.  
To keep notation simple, we suppress the index $\mu$  in what follows, and restore it only in the final results. 
 
Within second order perturbation theory, the density matrix factorizes as  $\rho(t)= \rho_S(t)\otimes\rho_E$, 
from the von Neumann equation one derives the integral equation\cite{breuer2002theory}, 
\beq{07}
\dot \rho_S(t) \approx \int_0^tds  \Gamma(s)[&&A(t-s)\rho_S(t-s)A(t)-\\ &&-A(t)A(t-s)\rho_S(t-s)]+h.c, \nonumber
\eeq
with $\Gamma (t)\equiv \langle B(t)B(0)\rangle_E$ the correlation function of the operator $B(t)$.

To proceed, one usually assumes that $\Gamma(s)$ is short ranged, and then makes a Markovian approximation,
 $\rho_S(t-s)\to \rho_S(t)$. (Keeping non-Markovian terms is discussed in Appendix A.) 
 Then, decomposing $A(t)$ into its eigenmodes of frequencies $\nu_{-j}=-\nu_j$ (and $A_j^\dagger=A_{-j}$)
\beq{09}
A(t)=\sum_{j=-J,...,J}A_j\eexp{-i\nu_jt}=A^\dagger(t),
\eeq
and taking the long time limit one obtains 
\beq{10}
\dot{\rho}_S= \sum_{j,k}\tilde\Gamma(\nu_j)\eexp{i(\nu_k-\nu_j)t}[A_j \rho_S A_k^\dagger-A_k^\dagger A_j\rho_S ]+h.c. \quad
\label{e10}
\eeq
with the couplings $\Gamma(\nu_j)$ defined as 
\beq{11}
\tilde\Gamma(\omega)=\int_0^\infty ds\Gamma(s)\eexp{i\omega s}=\half \gamma(\omega)+i\,\im \tilde\Gamma(\omega)\,.
\eeq
The generalized  Lindblad's equation  \eqref{e10} is the cornerstone of our analysis, what we examine beyond
the secular approximation.

\begin{table}[t!]
 \centering
\begin{tabular}{| c | c | c | c |}
 \hline
 Frequency &  $4/b_x \cdot A_{x,j} $ & $4/b_y \cdot A_{y,j} $   & $2/b_z \cdot A_{z,j} $  \\ \hline
$\nu_0=0$ &  $0$ & $0$ & $- \cos\theta \sigma_z$  \\ \hline
$\nu_1=\Omega'$ &  $0$ & $0$ & $\hphantom{-} \sin\theta \sigma_-$  \\ \hline
$\nu_2=\Omega'+\omega$ &  $-(1+\cos\theta)\sigma_-$ & $-i (1+\cos\theta)\sigma_-$ & $0$  \\ \hline
$\nu_3=\Omega'-\omega$ & $\hphantom{-} (1-\cos\theta)\sigma_-$ & $- i (1-\cos\theta)\sigma_-$  & $0$  \\ \hline
$\nu_4=\omega$ & $ \sin\theta \sigma_z$ & $- i  \sin\theta\sigma_z$  & $0$  \\ \hline
\end{tabular}
\caption{The operators appearing in Eq.{(\ref{e09})} in the XYZ coupling defined in Eq.(\ref{e15a}), $\sigma_{\pm}=(\sigma_x\pm i \sigma_y)/2$. In case of XXZ coupling the operators $A_{x,j}$ are given by the sum of $A_{x,j}$ and $A_{y,j}$ of the XYZ case, 
and similarly by $\displaystyle{\sum_{\mu=x,y,z}}A_{\mu,j}$ in the XXX coupling.}
\label{tab:Afourier}
\end{table}

 Usually \cite{shnirmanbook,schlosshauer,hauss,makhlin,shnirman2} one makes an additional assumption of keeping only secular terms
 with $\nu_j=\nu_k$, sometimes referred to as "\emph{modified rotating wave approximation}" \cite{HonePRE2009}.
 In this limit,  terms proportional to $\im\tilde\Gamma(\nu_j)$ just renormalize the  subsystem's Hamiltonian 
 (produce a Lamb shift) and can thus be dropped \cite{shnirmanbook}, and a usual Lindblad equation is recovered,  
\begin{align}
\dot{\rho}^{\rm sec}_S= 
\sum_{j}\gamma(\nu_j)\bigl\{ A_j\rho^{\rm sec}_S A_j^\dagger-\half A_j^\dagger A_j \rho^{\rm sec}_S - \half \rho^{\rm sec}_S A_j^\dagger A_j 
\bigr\} \;.
\label{e12}
\end{align}

To appreciate the role of the non-secular terms assume that an equilibrium solution $\rho_{eq}$ is found for the secular Eq. (\ref{e12}), and that deviations from equilibrium decay to it exponentially,  $\delta\rho(t) \sim \delta\rho(0) \exp(- \Gamma t)$.  Treating then the non-secular terms 
of Eq.~\eqref{e10} iteratively, one can immediately see that they generate corrections $\propto1/(i(\nu_k-\nu_j)+\Gamma)$, clearly 
demonstrating critical regions with  $|\nu_k-\nu_j|\lesssim \Gamma$.
In these regions the non-secular terms become important, and the secular approximation fails. 

Recovering the indices $\mu$ in Eq.~(\ref{e10}), the time evolution of the density matrix is given by
\begin{align} \label{eq:bloch_operator}
\dot{\rho}_S= \sum_{\mu,j,k}\tilde\Gamma_{\mu}(\nu_j)\eexp{i(\nu_k-\nu_j)t}[A_{\mu,j}& \rho_S A_{\mu,k}^\dagger-\\ \nonumber &-A_{\mu,k}^\dagger A_{\mu,j}\rho_S ]+h.c.,
\end{align}
where $\mu$ runs over statistically independent noise components, see Table \ref{tab:Afourier}.

\section{Applying the Lindblad equation to the edge state}

Let us now combine the results of the previous sections to investigate  the fate of the driven spin Hall system coupled to an environment. 
We start by deriving the time evolution operator for $\tilde{\cal H}_S(p)$. We note first that $\tilde{\cal H}_S(p)$ becomes static in the rotating frame, i.e. using the transformation $\eexp{\half i\omega t\sigma_z}$ that yields the Hamilitonian $\half(p-\omega)\sigma_z-\half\Omega\sigma_x$. Next we rotate into the $z$ axis by $\eexp{\half i\theta\sigma_y}$ where $\sin\theta=-\Omega/\Omega',\,\cos\theta=-\delta\omega/\Omega'$, leading to the Hamiltonian ${\cal H}^1_S(p)=\half\Omega'\sigma_z$. Finally, the time evolution w.r.t. ${\cal H}^1_S(p)$  is $\eexp{-\half i\Omega' t\sigma_z}$, hence the total evolution operator is
 \beq{33} \label{eq:U}
  U_S(t)=\eexp{-\half i\omega t\sigma_z} \eexp{- \half i\theta\sigma_y}  \eexp{-\half i\Omega't\sigma_z} 
 \eeq
We note that the conventional evolution operator is $U_S(t)\eexp{\half i\theta\sigma_y}$ (which is the identity at $t=0$). For either forms the interaction picture has  ${\cal H}_S(p)=0$, we find the form Eq.~\eqref{eq:U} to be more convenient. 
 
 Then we express $\tilde {\cal H}_{SE} $ in this rotated interaction picture as 
 \beq{19}
{\cal H}_{SE}=&&\sum_{j=\pm 1, 0}A_{z,j}\eexp{-i\nu_j t}Z(t)+\\&&+\sum_{j=\pm 2,\pm 3, \pm 4}A_{x,j}\eexp{-i\nu_j t}X(t)+A_{y,j}\eexp{-i\nu_j t}Y(t) \nonumber
\eeq
with the operators and the corresponding frequencies indicated in Table~\ref{tab:Afourier}. Having all operators 
$A_{\mu,j}$ at hand, we can now proceed and construct  the non-secular and secular Lindbald equations,  Eq.~\eqref{e10}. 
and Eq.~\eqref{e12}.

\subsection{Secular Lindblad equation}

We can apply the secular approximation  
in the limit, where all $\nu_j$ are sufficiently different relative to linewidths. Moreover, for infinitesimal system-bath coupling, the secular approximation becomes exact. This can be seen e.g. by noticing that rescaling time by 
$\alpha$ in Eq.~\eqref{eq:bloch_operator} upscales the frequencies of the non-secular oscillations.
It is useful to expand the density matrix $\rho_S(t)$ in Pauli matrices as 
\begin{equation}
\rho_S(t) = \half +  \sum_\mu \rho_\mu(t) \; \sigma_\mu. 
\end{equation}
In this basis 
the secular Lindblad equations read 
\beq{23}
\frac{d\rho_x}{dt}&=&-(\Gamma_\varphi^*+\half\Gamma_\downarrow+\half\Gamma_\uparrow)\rho_x\equiv -\frac{1}{T_2}\rho_x\nonumber\\
\frac{d\rho_y}{dt}&=& -\frac{1}{T_2}\rho_y\nonumber\\
\frac{d\rho_z}{dt}&=&-(\Gamma_\downarrow+\Gamma_\uparrow)(\rho_z-\rho_z^0)\equiv -\frac{1}{T_1}
(\rho_z-\rho_z^0)
\label{eq:bloch}
\eeq
with the equilibrium values $\half\langle\sigma_z\rangle_0=\rho_z^0=\half\frac{\Gamma_\uparrow
-\Gamma_\downarrow}{\Gamma_\uparrow+\Gamma_\downarrow},\,\rho_x^0=\rho_y^0=0$, 
and the emerging relaxation rates defined as:
\beq{22}
\Gamma_{\uparrow/\downarrow}&=&\frac{b_z^2}{4}\sin^2\theta\gamma_z(\mp\Omega')
\\
&+&\sum_{\mu\in(x,y)}\frac{b_\mu^2}{16}[c_{-}^2\gamma_\mu(\mp\Omega'\pm\omega)
+c_{+}^2\gamma_\mu(\mp \Omega'\mp \omega)]
\nonumber\\
\Gamma_\varphi^*&=&\frac{b_z^2}{2}\cos^2\theta\gamma_z(0)
+ \sum_{\mu\in(x,y)}\frac{b_\mu^2}{8}\sin^2\theta[\gamma_\mu(\omega)+
\gamma_\mu(-\omega)].
\nonumber
\eeq
Equations  \eqref{eq:bloch}  assume the form of  standard Bloch equations in the interaction picture with equilibrium in the $z$ direction. We note that $\Gamma_\downarrow,\Gamma_\uparrow$ can also be derived by a simpler Golden rule calculation \cite{makhlin,hauss}, 
 in agreement with our method. 
The results are the same for the XXZ and the XXX coupling, excepting that in the former we have to take $\gamma_x=\gamma_y$, 
and in the latter case $\gamma_x=\gamma_y=\gamma_z$.

The stationary value $\rho_z^0$ also gives the steady state occupation numbers of the eigenstates of the system Hamiltonian in 
the rotated frame $H_S^1(p)=\half \Omega' \sigma_z$. Its two eigenstates give also the nonequivalent Floquet states in the laboratory 
frame \cite{BukovAdvPhys2015}. The occupation of the state with  lower energy in the rotated frame is $n_{-}(p)=\half-\rho_z^0$. 

\begin{figure}
\centering
\includegraphics[width=8.4cm]{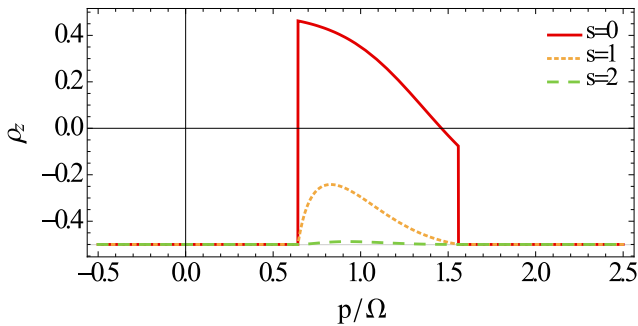}
\caption{The nonvanishing element of the density matrix in the stationary state at zero temperature ($\omega=1.1 \Omega$, $\alpha\rightarrow 0$ limit). The three curve correspond to different bath spectral functions $J(\nu)$. The excitations in the steady state are with stronger weight in the sub-Ohmic ($s=0$) case compared to Ohmic ($s=1$) or super-Ohmic environments ($s=2$).}
\label{fig:rhosecular}
\end{figure}

At zero temperature a sharp difference shows up between the occupation profiles in the cases of small frequency ($\omega<\Omega$) and large 
frequency driving, irrespective of the actual type of the bosonic heat bath. In the former case $\Gamma_{\uparrow}\equiv 0$, hence the steady state 
is described by filling the lowest lying states of $H_S^1(p)$. However, if $\omega>\Omega$, there is a narrow domain in the momentum space 
($p_{-}^*<p<p_+^{*}$, $p_{\pm}^*=\omega\pm\sqrt{\omega^2-\Omega^2}$), where $\Gamma_{\uparrow}\neq 0$, correspondingly the steady state contains 
excitations with respect to the rotating frame Hamiltonian. Depending on the spectral functions of the baths, inverse population is achieved in this region, see FIG.~\ref{fig:rhosecular}.

\begin{figure*}
\centering
\includegraphics[width=5.8cm]{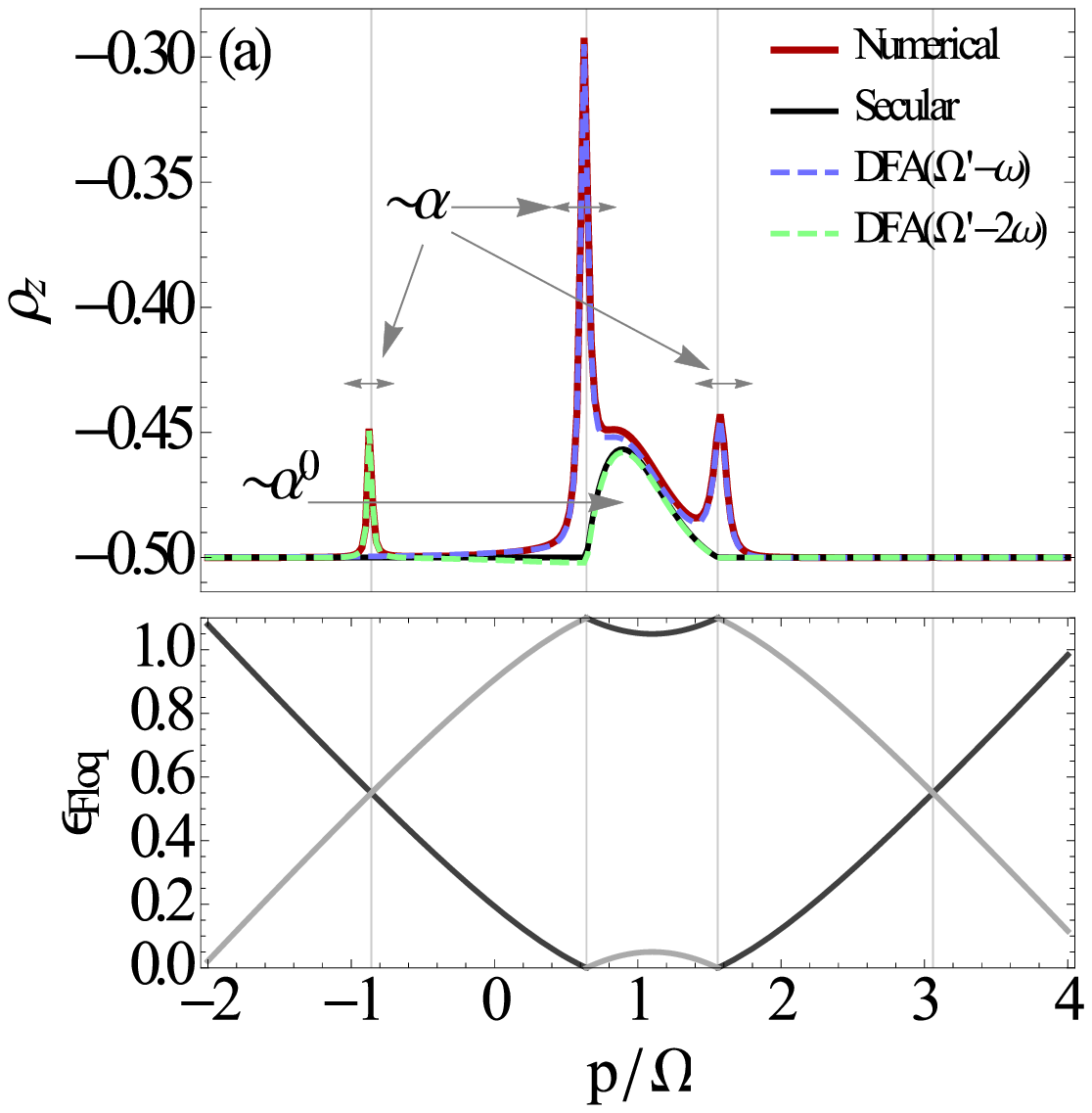}
\includegraphics[width=5.8cm]{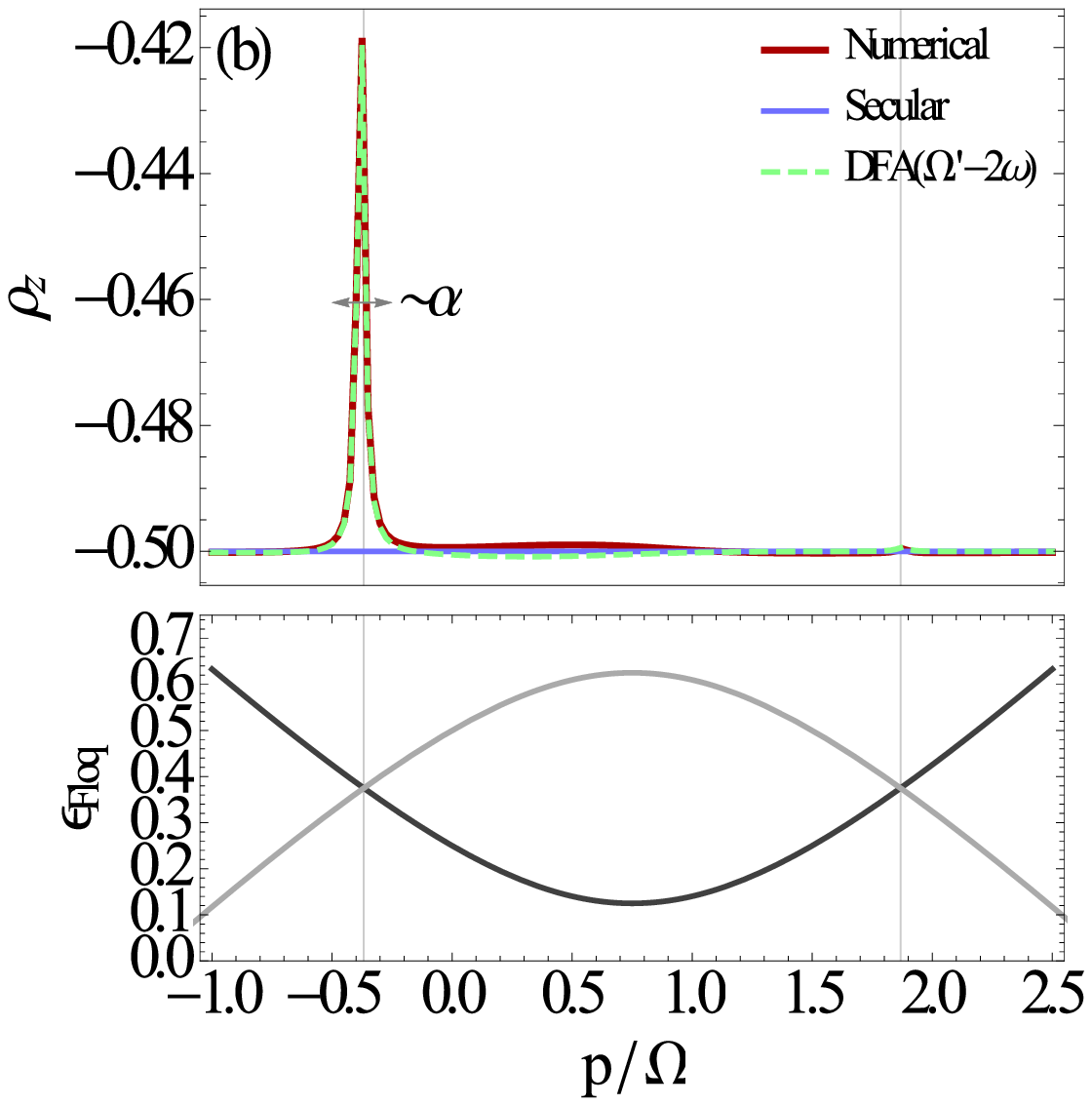}
\includegraphics[width=5.9cm]{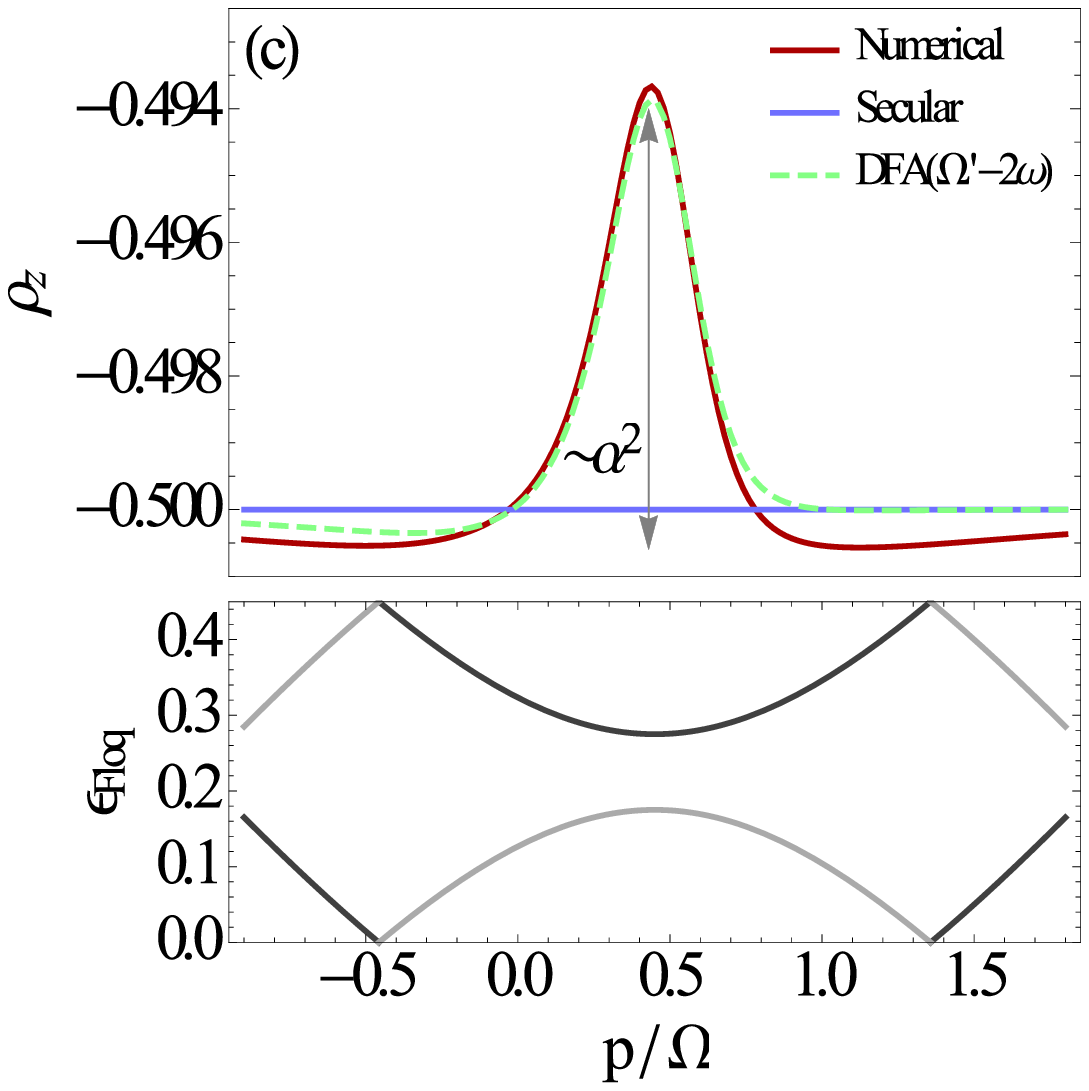}
\caption{Average stationary value of $\rho_z$ for various $\omega/\Omega$ ratios in the XXX case ($s=1$, $\alpha=0.05$).
(a) $\omega>\Omega$, correspondingly $\rho_z$ deviates from the secular solution at critical momenta $p_{\pm}^{*}$ ($\Omega'-\omega\approx 0$) 
and $p_{\pm}^{**}$ ($\Omega'-2\omega\approx 0$). This is attributed to two-photon transitions, where the Floquet bands touch each other (lower panel). 
(b) At lower frequency ($\Omega>\omega>\half\Omega$) only one dangerous non-secular term survives, giving rise to peaks at $p_{\pm}^{**}$. 
(c) When $\omega\lesssim\half\Omega$ the secular approximation still gets corrections because the dangerous frequency $\Omega'-2\omega$ is small 
at $p\approx \omega$. The actual values of $\omega/\Omega$ are $1.1$ in (a), $0.75$ in (b) and $0.48$ in (c).}
\label{fig:rhofloq}
\end{figure*}

\subsection{Beyond the secular approximation}

The Bloch equations are rewritten as
\begin{align}  \label{eq:Bloch}
 \frac{\dd\vrho(t)}{\dd t}=\vB(t)\vrho(t)+\vb(t),
\end{align}
where
\begin{align} \label{eq:bBfour}
 \vB(t)&=2\sum_{\mu,j,k}  \tilde{\Gamma}_\mu(\nu_j) \eexp{i (\nu_k-\nu_j)t}[\va_{\mu,j}\va_{\mu,k}^{+}- \\ &\hspace{4cm}-\mathbb{I}\,\va_{\mu,j}\cdot\va_{\mu,k}^{*}]+h.c.\nonumber\\
 \vb(t)&= \sum_{\mu,j,k} i \tilde{\Gamma}_\mu(\nu_j) \eexp{i (\nu_k-\nu_j)t} [\va_{\mu,j}\times\va_{\mu,k}^{*}]+h.c.
\end{align}
with $A_{\mu}(t)=\va_{\mu}(t)\cdot \vsigma$ and $\va_\mu(t)=\sum_j \va_{\mu,j} \eexp{-i \nu_j t}$, and $\va_{\mu,j}$ can be identified from TABLE~\ref{tab:Afourier}.
In the secular approximation only the $j=k$ terms are kept, i.e. only $\tvB(0)$ and $\tvb(0)$.  
The frequencies appearing in the above expansions in the XYZ and XXZ cases are $\pm$(0, $\Omega'$, $2\omega$, $\Omega'\pm 2\omega$) 
in $\vb(t)$, and $\pm$(0, $\Omega'$, $2\Omega'$, $2\omega$, $\Omega'\pm 2\omega$, $2(\Omega'\pm\omega)$) in $\vB(t)$. 
In the XXX case, additional frequencies $\pm(\omega, \Omega' \pm \omega)$ to $\vb(t)$ and 
$\pm(\omega,\Omega'\pm\omega,2\Omega'\pm\omega)$ to $\vB(t)$ appear.
For the full solution of the problem, all these terms should be taken into account, which is easy to implement numerically. 
Generally, all the above Fourier components appear in the time evolution of the density matrix, $\rho_x$ and $\rho_y$ 
oscillate around $0$, while $\rho_z$ oscillates around a finite stationary value. The secular approximation works well 
if none of these frequencies are close to zero.
In the case when one of these frequencies nearly vanish, the stationary values are tuned away from the secular ones. 
The possibly dangerous terms that can vanish at certain momenta, possess frequencies as $\nu_{*}=\Omega'-\omega$ and $\nu_{**}=\Omega'-2\omega$. ~\footnote{In principle 
$2\Omega'-\omega$ could be dangerous as well, but it becomes unimportant due to vanishing matrix elements.} 
When these frequencies become small, the deviation from the secular approximation grows, which appears as a peak in the stationary components of $\vrho(p)$. 

In the vicinity of these points, analytical solutions are possible within the Dominant Frequency Approximation (DFA).
When the frequency of some non-secular terms approaches zero, it drives the
solution away from the secular one. The single smallest frequency 
appearing among the non-secular terms is the dominant one. Keeping this single frequency, Eq. \eqref{eq:bloch_operator} can be transformed to a time independent equation, that is readily solved, as detailed in the Appendix.

The full numerical solution of the Eq.~\eqref{eq:Bloch}, together with various approximate results are shown in  Fig.~\ref{fig:rhofloq}, visualizing 
the momentum dependence of the average value of $\rho_z$ in the Ohmic case. Note that $\rho_z$ can become smaller than $-1/2$, which is a 
common feature in other non-secular approaches as well\cite{kamleitner}.
The secular approximation clearly breaks down at certain momenta, and is outperformed by the DFA there.
~\footnote{In the secular approximation only the real part of $\tilde{\Gamma}(\nu)$ appears in the stationary solution, correspondingly it behaves well for any values of $s$, even for the white noise limiting case ($s=0$) of the 
sub-Ohmic regime. However, if non-secular terms are considered, one cannot neglect the imaginary parts. If $s=0$, Im$\tilde{\Gamma}(\nu)$ diverges logarithmically for small frequency, which also implies unphysical 
stationary states with diverging components of the density matrix. This divergence originates from the extension of the upper limit of the integral in the Bloch-Redfield equation to $\infty$ in Eq.~(\ref{e07}). Without 
extending the integral to $\infty$, the equation can be used to study the short time dynamics in the $s=0$ case, but it cannot describe the stationary state.}

Despite the several Fourier coefficients appearing in $\vrho(t)$, the spins exhibit periodic oscillations in the laboratory frame with frequency $\omega$ and higher harmonics. Indeed, switching back to the Schr{\"o}dinger picture ($\tilde{\vrho}(t)$) after applying the Markovian approximation on Eq.~\eqref{e07} results in a differential equation, which only involves frequencies $0$, $\omega$ and its higher harmonics ($\Omega'$ affects only coefficients via the $s$ integration). This is favorable for numerical calculations but is also disadvantageous for analytical treatment compared to the interaction picture, which gives a natural ground to investigate the resonances and provides approximate solutions for the time evolution of the density matrix.

\section{Photocurrent along the edge} \label{sec:edgecurr}

Armed with the knowledge of the density matrix, we now focus on measurable quantities. 
Due to the electromagnetic field, a  net electric current and magnetization due to the magnetoelectric effect\cite{hasankane} is induced along the edge. 
Without the environment\cite{dorafloquet}, this current was found to be quantized based in the average energy concept in the adiabatic limit, giving way to
dissipative charge transport through a topological transition with increasing frequency $\omega$. 
The photocurrent along the edge of a spin-Hall insulator in the laboratory frame is determined as $j=-e\int \frac{\dd p}{2\pi} \tr\{\tilde{\rho}(t)\half\sigma_z\}$, which is expressed by the components of $\vrho$ in the interaction picture as 
\begin{align} \label{eq:current}
j= e\int_{-\Lambda}^{\Lambda}\frac{\dd p}{2\pi} \, \frac{\delta\omega}{\Omega'}\rho_z-\frac{\Omega}{\Omega'}\left(\rho_x\cos\Omega' t-\rho_y\sin\Omega' t\right) \,,
\end{align}
which inherits the $2\pi/\omega$ periodicity from $\tilde{\vrho}(t)$. In the secular approximation only the DC component survives, as $\rho_{x,y}=0$ and $\rho_z$ is time independent. In the low frequency limit ($\omega<\Omega$) $\rho_z\equiv -\half$, yielding 
$j_c=\frac{e\omega}{2\pi}$, which we call the quantized value following Ref.~\cite{dorafloquet}. When $\omega>\Omega$, there are regions where 
$\rho_z$ deviates from $-\frac{1}{2}$ (Figs.~\ref{fig:rhosecular},\ref{fig:rhofloq}), implying the breakdown of the quantization. Near the 
critical point the deviation form the quantized current is $\Delta j=j-j_c\sim (\omega-\Omega)^{s+3/2}$, with the exponent depending on the low 
frequency asymptotics of the bath spectral function. 

\begin{figure}
\centering
\includegraphics[width=8.4cm]{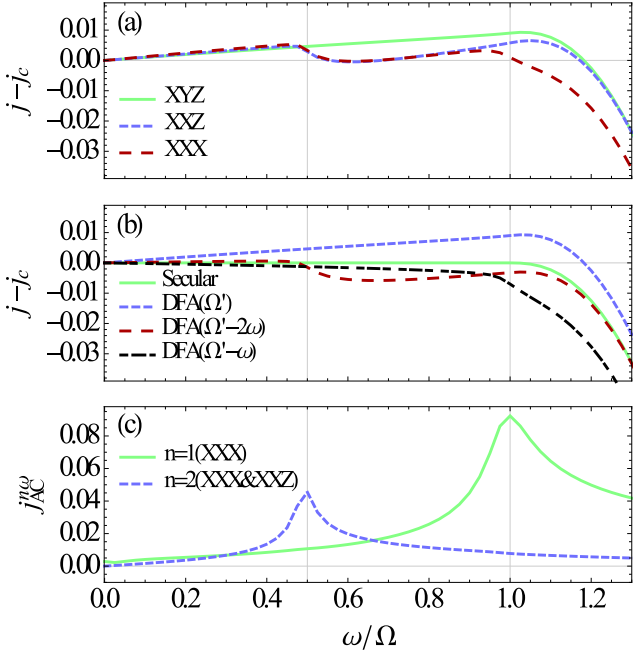}
\caption{Stationary current induced by a circular driving 
on a QSH edge coupled to an Ohmic bath (the vertical units 
are in $j_c(\omega=\Omega)=\frac{e\Omega}{2\pi}$, $\alpha=0.1$). (a) Numerical solution. 
(b) Analytical results for the symmetric case with $b_x=b_y=b_z$, $\Gamma_x=\Gamma_y=\Gamma_z$.
The secular approximation
corresponds to an infinitesimal system-bath coupling, where 
the quantization is exact until the driving frequency reaches the Rabi frequency. 
The 2 photon processes at finite coupling constants violate the quantization, which 
become effective at $\omega\sim \half\Omega$. This is well captured in the DFA. 
(c) The photon-resonances also give rise to an AC current with frequency $n\omega$, $n\in\mathbb{N}$.}
\label{fig:currall}
\end{figure}

The current obtains $\sim \alpha$ corrections to the secular approximation due to the non-secular terms, and the quantization of the current ceases to be exact at finite system-bath couplings (FIG.~\ref{fig:currall}(a,b)). The corrections have dual origin. On the one hand, due to photon absorption resonances near $\omega\approx \half \Omega',\Omega'$, $\rho_z$ deviates from $-\half$ even for $\omega<\Omega$. On the other hand, the $x,y$ components of the density matrix acquire oscillations at frequency $\Omega'$, which also contribute to the DC current in Eq.~\eqref{eq:current}. The resonances produce a sharp breakdown in the current at $\omega\approx \half \Omega$ due to a dangerous non-secular term present in the XXZ and XXX cases. 
This behavior is captured in the DFA. FIG.~\ref{fig:currall}(b) shows the crossover at $\omega\sim\half\Omega$ due to the $\Omega'-2\omega\approx 0$ resonance, and also the effect of the $\Omega'-\omega\approx0$ resonance on the DC current. The DFA with frequencies $\Omega'-\omega$ and $\Omega'-2\omega$ cannot yield nonzero $\Omega'$ fluctuations. To describe the effect of the second term in Eq.~\eqref{eq:current}, one has to study the DFA with frequency $\Omega'$ (FIG.\ref{fig:currall}(b)). 
The deviations of the DFA with different frequencies compared to the secular approximation are approximately additive, and one can combine them to achieve a good approximation for the total DC current.

At finite system-bath coupling, in addition to the DC component, the stationary current is also characterized by nonvanishing AC contributions, showing peaks as a function of $\omega$ (FIG.~\ref{fig:currall}(c)). The $\omega$ frequency Fourier component of the current originates from the $\omega$, $\Omega'\pm \omega$ Fourier terms in $\vrho(t)$, which are present only in the XXX coupling. The $2\omega$ and $\Omega'\pm 2\omega$ components of $\vrho(t)$ are responsible for the $2\omega$ harmonic AC current, which therefore is present in the XXZ and XXX couplings. The XYZ coupling does not show any alternating current.

In general, finite temperature also breaks down the quantization of the current. However, in the special case of $b_x=b_y=0$, 
when the occupation is thermal, the current remains quantized even at finite temperature. 
It is also worth mentioning that in this case there are no critical points at all (see e.g. $\rho_z^0$ together with the definitions of $\Gamma_{\uparrow,\downarrow}$, and also TABLE~\ref{tab:Afourier}), and the quantization remains valid for all frequencies. 

\begin{figure}
\centering
\includegraphics[width=8.4cm]{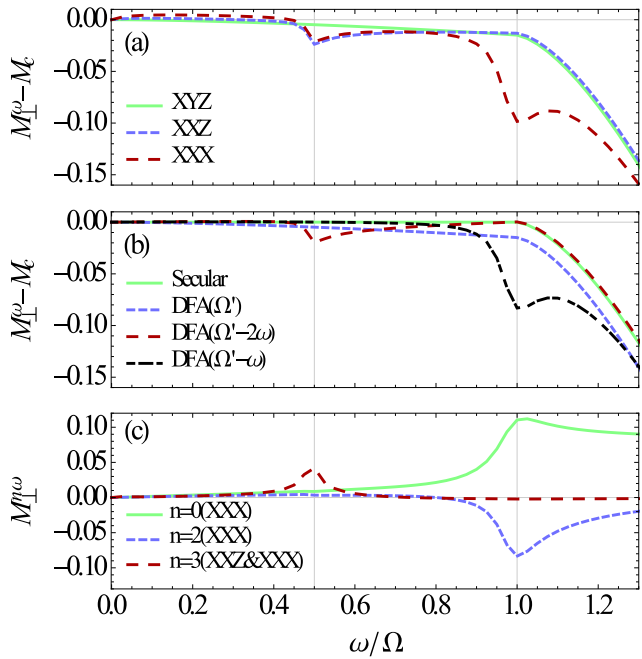}
\caption{Frequency dependence and crossovers in the transverse magnetization (in units of $\frac{\Omega}{2\pi}$). In contrast to the current, the transverse magnetization oscillates with frequency $\omega$ in the $\alpha\rightarrow 0$ limit, with amplitude $M_c$ for $\omega<\Omega$. (a) Deviation from the limiting value $M_c$ at finite system-bath coupling, numerical solution ($\alpha=0.1$) (b) DFA with the two dangerous frequencies and with $\Omega'$ is capable to reconstruct all the features in the numerical solution. (c) Other Fourier components of the magnetization. }
\label{fig:magnall}
\end{figure}

The integrated expectation value of $\sigma_z$ determines both the current and the $z$ component of the edge magnetization. The magnetization in the $xy$ plane is calculated similarly, 
and, as in Ref.~\cite{dorafloquet}, it exhibits a circular motion on average with frequency $\omega$: 
\begin{align}
M_{\perp}^{\omega}&=\int_{-\Lambda}^{\Lambda}\frac{\dd p}{2\pi} \, \frac{1}{2} \expv{\sigma_x \cos(\omega t)+\sigma_y \sin(\omega t)}\\
&=-\int\frac{\dd p}{2\pi} \frac{\Omega}{\Omega'} \rho_z+\frac{\delta\omega}{\Omega'}(\rho_x\cos\Omega't-\rho_y\sin\Omega't)
\end{align}
Similar to the current, only the first term survives in the secular approximation, and in further analogy to the quantized current we define $M_c=\int\frac{\dd p}{2\pi} \frac{\Omega}{\Omega'}\half=\frac{\Omega}{2\pi}\log \frac{2\Lambda}{\Omega}$, which is independent of $\omega$ and logarithmically divergent in the cutoff parameter $\Lambda$. The crossovers in $\rho_z$ as a function of $\omega$ are also revealed in transverse magnetization, which can be highlighted by subtracting the low frequency transverse magnetization $M_c$ as a reference value (FIG.~\ref{fig:magnall}). 
In the XXX and XXZ cases the magnetization acquires a finite $3\omega$ component due to the $\Omega'\pm 2\omega$ components of $\vrho$, and the $\Omega'-\omega$ resonance in the XXX case gives rise to a finite static magnetization in the $xy$ plane together with the second harmonic (FIG.~\ref{fig:magnall}(c)).

\section{Conclusion}

We have investigated the fate of a spin-Hall edge state, coupled to dissipative environment,  in the presence of circularly polarized electromagnetic field.
Without the environment, the Floquet solution of the problem features an electromagnetic field induced photocurrent, being quantized in the adiabatic regime\cite{thouless}
 and crossing
over to dissipative charge transport with increasing frequency\cite{dorafloquet}.
These results were obtained using the average energy concept for the occupation of the Floquet steady state.
In the presence of dissipation, the filling of the Floquet states is determined by a Lindblad equation, which we investigate both analytically and numerically.
The photocurrent is only quantized in the strict adiabatic limit in the presence of finite bath coupling. 
 Nevertheless, deviations from perfect quantization in the adiabatic regime are tiny, as shown in Fig. \ref{fig:curravg}, and perfect quantization
is recovered in the limit of vanishing coupling to the environment.
With increasing frequency, deviations grow and the photocurrent becomes dissipative in nature. Our results show that couplings to environments are essential for treating Floquet systems and that these can be identified by measuring DC as well as AC observables of the system.

\begin{acknowledgments}
This research has been supported by the Hungarian Scientific Research Fund No. K101244,  K105149, K108676 and by the Bolyai program of the
Hungarian Academy of Sciences. One of us (BH) thanks A. Shnirman for valuable discussions and acknowledges support by the DFG through the DIP program (FO 703/2-1).
\end{acknowledgments}

\bibliographystyle{apsrev}

\bibliography{refgraph}

\vspace{2cm}
\clearpage
\onecolumngrid

\appendix

 \section{Non-Markovian equation}

\begin{figure}[h!]
\centering
\includegraphics[width=8.4cm]{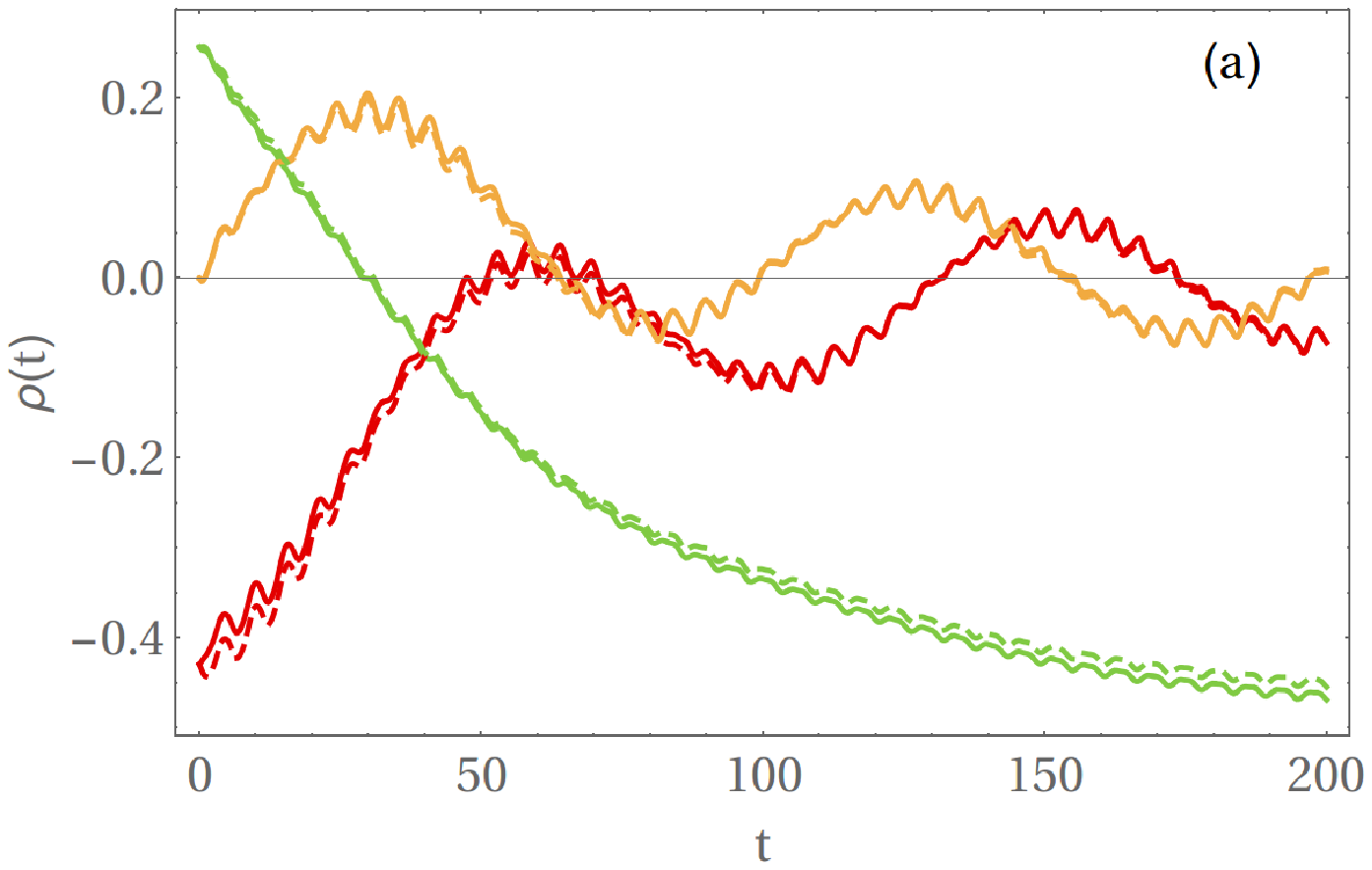}
\includegraphics[width=8.4cm]{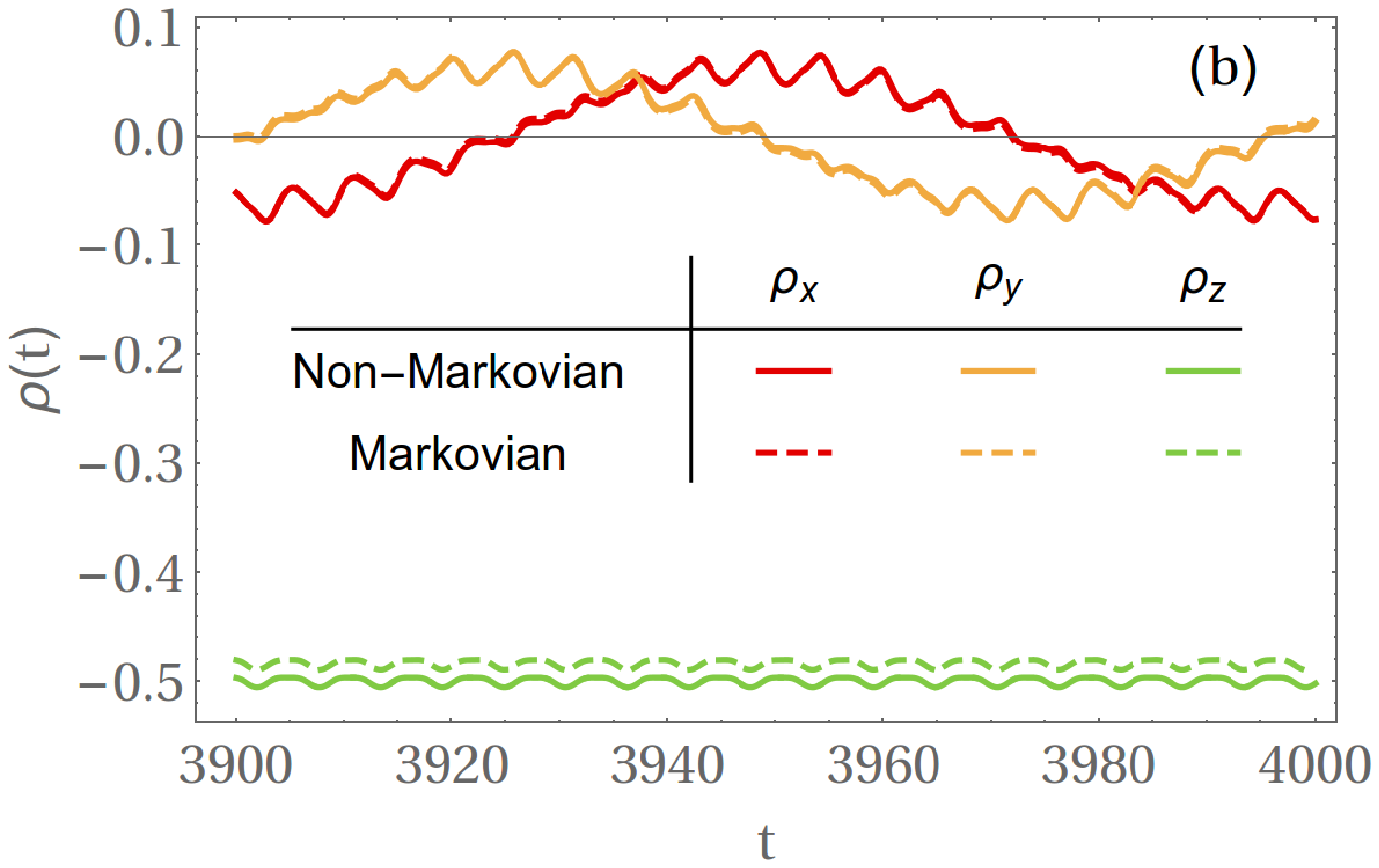}
\caption{Comparison of the Markovian and non-Markovian time evolution in an Ohmic environment for short (a) and long (b) times.}
\label{fig:nonmark}
\end{figure}

If we drop the Markovian assumption, have to solve the following integro-differential equation,
\begin{align} \label{eq:rhodotNMV}
 \frac{\dd}{\dd t}\hrho(t)=\sum_{\mu\in \{x,y,z\}}\int_{0}^{t} \dd s\, \Gamma_{\mu}(s) [\hA_{\mu}(t-s)\hrho(t-s)\hA_{\mu}(t)-\hA_{\mu}(t)\hA_{\mu}(t-s)\hrho(t-s)]+h.c.
\end{align}
or after expanding $\hrho(t)=\frac{1}{2}+\vrho(t)\cdot \vsigma$ and $\hA_{\mu}(t)=\va_{\mu}(t)\cdot \vsigma$,
\begin{align}  \label{eq:BlochNMV}
 \frac{\dd\vrho(t)}{\dd t}=\int_0^t \dd s \, \vB(t,s) \vrho(s)+\vb(t)
\end{align}
where
\begin{align}
 \vB(t,s)&=4\Re \{\sum_{\mu} \Gamma_{\mu}(s) [\va_{\mu}(t-s)\cdot\va_{\mu}^{T}(t)-\mathbb{I}\,\va_{\mu}(t-s)\cdot\va_{\mu}(t)]\}\\
 \vb(t)&=2\Re \{\sum_{\mu} \int_{0}^{t} \dd s\,  \Gamma_{\mu}(s) i[\va_{\mu}(t-s)\times\va_{\mu}(t)]\}
\end{align}
To test the Markovian approximation, we solved numerically the above integro-differential equation with Heun's method (a two-stage predictor-corrector method), and compared the solution with the Markovian approximation (FIG.~\ref{fig:nonmark}). There is a very small quantitative difference in the stationary states, but the qualitative picture does not change.

\section{Dominant frequency approximation (DFA)} \label{sec:DFA}

We observe that the operators $A_{\mu,j}$ in TABLE~\ref{tab:Afourier} are either proportional to $\sigma_{\pm}$
or to $\sigma_z$, thus they obtain only a phase factor under a rotation around the $z$ axis. The secular terms
always consist of an operator $A_{\mu,j}$ and its hermitian conjugate, hence they do not transform under the
rotation $U=\eexp{\half i \nu t \sigma_z}$. On the other hand, the non-secular terms acquire a phase factor,
which allow us to transform the dominant frequency term to be time independent, without spoiling the secular terms.

Let $\nu_d\in\{\nu_{*},\nu_{**}\}$ denote the dominant frequency (the method works for the frequencies $\Omega'$,
$\half\Omega'-\omega$ as well), and $\vrho'(t)=R(\nu_d t)\vrho(t)$ is the vector representation of the density matrix
after the rotation, where $R(\nu t)$ is the $3\times 3$ rotation matrix around the $z$ axis in the positive direction.
Keeping only the time independent terms in the Bloch equation for ${\vrho}'(t)$  yields $\dot{\vrho}'(t)=\tvB_d \vrho'(t)+\tvb_d$,
where $\tvB_d=\sum \tvB(\nu)-R\dot{R}^{-1}$ and $\tvb_d=\sum \tvb(\nu)$ with the sum going through $\nu\in V_d=\{0,\pm\nu_d,\pm 2\nu_d\}$.
The matrix structure of the Fourier components are shown in TABLE~\ref{tab:Bfourier}, the matrix elements are given explicitly in the Appendix~\ref{sec:app_matrixelem}.
\begin{table}[h!]
 \centering
\begin{tabular}{| c  | c  | c | }
 \hline
$ \hspace{2em} \tvB(0)$ \hspace{1em} $\tvb(0)$ &  \hspace{1.5em} $\tvB(\nu)$  \hspace{1.5em} $\tvb(\nu)$ & \hspace{2em}$\tvB(2\nu)$ \hspace{0.1em}  $\tvb(2\nu)$   \\ \hline
$\begin{pmatrix}
  \eta & \eta_2 & 0 \\
  -\eta_2 & \eta & 0 \\
  0 & 0 & \xi
 \end{pmatrix} $ \hspace{-1em} $\begin{pmatrix} 0\\0\\ \epsilon\end{pmatrix}$ &
$\begin{pmatrix}
  0 & 0 & \beta_1 \\
  0 & 0 & i \beta_1 \\
  \beta_2 & i \beta_2 & 0
 \end{pmatrix}$  \hspace{-1em} $\begin{pmatrix} \chi\\i \chi\\ 0\end{pmatrix}$ &
$\begin{pmatrix}
  \delta & i \delta & 0 \\
  i \delta & -\delta & 0 \\
  0 & 0 & 0
 \end{pmatrix}$ \hspace{-1em} $\begin{pmatrix} 0\\0\\ 0\end{pmatrix}$  \\ \hline
\end{tabular}
\caption{Matrix structure of the various Fourier components ($\nu\in\{\Omega'-\omega,\Omega'-2\omega, \Omega', \half\Omega'-\omega\}$)
appearing in Eq.~\eqref{eq:bBfour}. All the matrix elements are $\sim \alpha$, $\epsilon=-\frac{1}{2}(\Gamma_{\uparrow}-\Gamma_{\downarrow})$,
$\xi=\Gamma_{\uparrow}+\Gamma_{\downarrow}$ and $\eta=\frac{1}{2}(\Gamma_{\uparrow}+\Gamma_{\downarrow})+\Gamma_{\phi}^{*}$. The Lamb shift $\eta_2$ together with the other matrix elements are listed in Appendix~\ref{sec:app_matrixelem}.}
\label{tab:Bfourier}
\end{table}

\begin{figure}[h!]
\centering
\includegraphics[width=8.4cm]{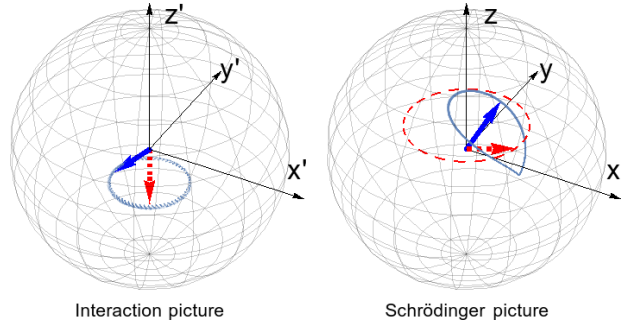}
\caption{Stationary state dynamics of the density matrix on the Bloch sphere in the interaction and Schr\"{o}dinger pictures. The curves show
the path of $\vrho$ close to a critical point, where the non-secular terms are non negligible. In the secular approximation (red dashed
curve and arrow) $\vrho$ reaches a constant value in the interaction picture, which corresponds to a circular motion in the laboratory frame (i.e. in the Schr\"{o}dinger picture). In the DFA $\vrho$ draws a circle around the secular solution in the interaction picture. This circle also has a fine structure, if we go beyond
the DFA (blue solid curve and arrow).}
\label{fig:blochrotlab}
\end{figure}

At this level of approximation $\vrho'$ achieves a constant stationary value determined by the matrix elements of the Fourier components. Going back to the interaction picture, we see a constant $\rho_z^{\text{stac}}$ and oscillating $\rho_{x,y}^{\text{stac}}$ with
$\frac{\pi}{2}$ phase difference between them (FIG.~\ref{fig:blochrotlab}). The amplitude of this oscillation is given by $\rho_{\perp}=\sqrt{\rho_x'^2+\rho_y'^2}$.
These steady state values are expressed as:
\begin{align} \label{eq:rhostactr2}
 \rho_z'^{\text{stac}}&=-\frac{\epsilon \zeta_1-\zeta_2}{\xi \zeta_1-\zeta_3} \qquad \rho_{\perp}'^{\text{stac}}=2\left| \frac{\zeta_4}{\xi \zeta_1-\zeta_3} \right|\\
 \zeta_1&=\eta^2-4|\delta|^2+(\nu-\eta_2)^2\\
 \zeta_2&=4 \Re\{\chi \beta_2^{*}(\eta+i(\nu-\eta_2))-2\chi\beta_2\delta^{*}\}\\
 \zeta_3&=4 \Re\{\beta_1 \beta_2^{*}(\eta+i(\nu-\eta_2))-2\beta_1\beta_2\delta^{*}\}\\
 \zeta_4&=(\chi \xi -\beta_1 \epsilon)(\eta+i(\nu-\eta_2))+\\&\hphantom{=}+2\beta_2(\chi^{*}\beta_1-\chi \beta_1^{*})+2\delta(\beta_1^{*}\epsilon-\chi^{*}\xi)
\end{align}
This expression makes it clear that as the dissipation strength $\alpha$ tends to zero, the solution approaches the secular one,
$\rho_z^{\text{stac}}=-\frac{\epsilon}{\xi}$, $\rho_{\perp}^{\text{stac}}=0$. This can be seen by observing that the only terms of
order $\alpha$ are $\epsilon \nu^2$ in the nominator of $\rho_z^{\text{stac}}$ and $\eta \nu^2$ in the denominators; all the others are
at least $\mathcal{O}(\alpha^2)$.  On the other hand, for any finite system-bath coupling strength, approaching the critical points
- where the dominant frequency vanishes - close enough, the secular approximation breaks down. Now we discuss separately the results of the DFA for the various couplings.

In the \emph{$U(1)$ symmetric XYZ case} there are no dangerous non-secular terms, and the full time evolution can be mapped to be exactly
time independent applying the method described above with $\nu_d=\Omega'$. This yields an analytical solution for the stationary fermion occupations.

The \emph{$U(1)$ breaking XYZ and XXZ case} has a single dangerous non-secular term corresponding to the frequency $\nu_{**}$
(because $\nu_{*}$ does not enter in $\vB(t)$ and $\vb(t)$). This vanishes if $\omega>\half\Omega$ at the critical momenta
$p_{\pm}^{**}=\omega\pm\sqrt{4\omega^2-\Omega^2}$, giving rise to peaks in the stationary values of $\rho_z$ and $\rho_{\perp}$.
In contrast to the singularity in the secular solution at $\omega=\Omega$, these peaks grow up gradually as $\omega$ is increased,
and are also present in the $\omega<\half\Omega$ case (see FIG.~\ref{fig:rhofloq}(c) for the same phenomenon in XXX case). We note
that Eq.~\eqref{eq:rhostactr2} is in the most general form, and it simplifies for $\nu_d=\nu_{**}$ as $\delta \equiv 0$ in this case.

In the \emph{XXX coupling}, in addition to $\nu_{**}$, the frequency $\nu_{*}$ becomes dangerous as well, and it vanishes at momenta
$p_{\pm}^{*}$ for $\omega>\Omega$. These are the momenta between which excitations are present in the steady state even in the secular
approximation. The contribution of the dominant frequency terms is that additional peaks grow at $p_{\pm}^*$ on the top of the secular
solution of $\rho_z$ and $\rho_{\perp}$ (FIG.~\ref{fig:rhofloq}(a)). In principle the frequency $\nu_{***}=\Omega'-\half\omega$ could be
dangerous as well, but the matrix elements $\chi=\beta_{1,2}\equiv 0$ at this frequency. Because of the vanishing matrix elements, it
does not change the secular behavior, see e.g. Eq.~\eqref{eq:rhostactr2}.

An equivalent way to look at the DFA is to take the Fourier transform of Eq.~\eqref{eq:Bloch}, which maps the differential equation to
 an (infinite) set of coupled linear equations $ i \tvrho(\omega_i)\omega_i=\sum_j \tvB(\omega_i-\omega_j)\tvrho(\omega_j)+\tvb(\omega_i) $.
The approximation is to keep only the dominant frequency $\nu_d$ in the expansion of $\tvrho$, neglecting the higher harmonics as well.
The generalization to keep more (dominant) frequencies is straightforward in this language, but analytically hardly treatable.

\section{Matrix elements} \label{sec:app_matrixelem}
The diagonal matrix elements of $\tvB(0)$ and $\tvb(0)$ were already given in the main text. The Lamb shift $\eta_2$ is 
\begin{scriptsize}
\begin{align}
 \eta_2= -i \frac{b_z^2}{8} (\frac{\Omega}{\Omega'})^2 &\left[ \Gamma_z^{*}(-\Omega')-\Gamma_z(-\Omega')-\Gamma_z^{*}(\Omega')+\Gamma_z(\Omega')  \right] +i\frac{b_x^2}{32}  \left\{ (1+\frac{\omega-p}{\Omega'})^2 \left[ \Gamma_x^{*}(\omega-\Omega')-\Gamma_x(\omega-\Omega')-\Gamma_x^{*}(\Omega'-\omega)+\Gamma_x(\Omega'-\omega) \right] \right.   \nonumber \\  
 & \left. +  (1-\frac{\omega-p}{\Omega'})^2 \left[ \Gamma_x^{*}(-\omega-\Omega')-\Gamma_x(-\omega-\Omega')-\Gamma_x^{*}(\Omega'+\omega)+\Gamma_x(\Omega'+\omega) \right]  \right\}  + "x\leftrightarrow y"
\end{align}
\end{scriptsize}
for the XYZ case, and the same for the XXZ/XXX cases are given by the substitution $\Gamma_y=\Gamma_x$, $\Gamma_y=\Gamma_z=\Gamma_x$.

The Fourier component $\nu_{*}=\Omega'-\omega$ appears only in the XXX case, with the matrix elements
\begin{scriptsize}
\begin{align}
 \chi&=\frac{(b_x-i b_y)b_z}{16}\left\{ \frac{\omega-p}{\Omega'}(1+\frac{\omega-p}{\Omega'})\left[\Gamma_x(\Omega'-\omega)-\Gamma_x^{*}(\omega-\Omega')+\Gamma_x^{*}(0)-\Gamma_x(0)\right]
 - \frac{\Omega^2}{\Omega'^2}\left[\Gamma_x^{*}(\omega)-\Gamma_x(-\omega)+\Gamma_x^{*}(\Omega')-\Gamma_x(-\Omega')\right] \right\}\\
 \beta_1&=\frac{(b_x-i b_y)b_z}{8}\left\{ \frac{\omega-p}{\Omega'}(1+\frac{\omega-p}{\Omega'})\left[\Gamma_x(\Omega'-\omega)+\Gamma_x^{*}(\omega-\Omega')\right]
 - \frac{\Omega^2}{\Omega'^2}\left[\Gamma_x(\Omega')+\Gamma_x^{*}(-\Omega')\right] \right\} \\
 \beta_2&=\frac{(b_x-i b_y)b_z}{8}\left\{ \frac{\omega-p}{\Omega'}(1+\frac{\omega-p}{\Omega'})\left[ \Gamma_x^{*}(0)+\Gamma_x(0)\right]
 - \frac{\Omega^2}{\Omega'^2}\left[\Gamma_x^{*}(\omega)+\Gamma_x(-\omega)\right] \right\}\\
 \delta&=\frac{(b_x-i b_y)^2}{32}(1+\frac{\omega-p}{\Omega'})^2\left[\Gamma_x(\Omega'-\omega)+\Gamma_x^{*}(\omega-\Omega') \right]
 \end{align}
\end{scriptsize}
but the second harmonic $2\nu_{*}$ is present in the XXZ and XYZ cases as well. In the former $\delta$ is identical to that of the XXX case, while for the latter
\begin{scriptsize}
\begin{align}
 \delta&=\frac{b_x^2}{32}(1+\frac{\omega-p}{\Omega'})^2\left[\Gamma_x(\Omega'-\omega)+\Gamma_x^{*}(\omega-\Omega') \right] - "x\leftrightarrow y"
 \end{align}
\end{scriptsize}

In the case of Fourier component $\nu_{**}=\Omega'-2\omega$ the second harmonic $\delta\equiv 0$ in all the coupling schemes. The other matrix elements are
\begin{scriptsize}
\begin{align}
 \chi&=\frac{b_x^2}{32}\frac{\Omega}{\Omega'}(1+\frac{\omega-p}{\Omega'})\left[\Gamma_x^{*}(\omega)-\Gamma_x(-\omega)+\Gamma_x(\Omega'-\omega)-\Gamma_x^{*}(\omega-\Omega') \right] -"x\leftrightarrow y"\\
 \beta_1&=\frac{b_x^2}{16}\frac{\Omega}{\Omega'}(1+\frac{\omega-p}{\Omega'})\left[\Gamma_x(\Omega'-\omega)+\Gamma_x^{*}(\omega-\Omega') \right]- "x\leftrightarrow y"\\
 \beta_2&=\frac{b_x^2}{16}\frac{\Omega}{\Omega'}(1+\frac{\omega-p}{\Omega'})\left[\Gamma_x^{*}(\omega)+\Gamma_x(-\omega) \right] - "x\leftrightarrow y"
 \end{align}
\end{scriptsize}
for the XYZ case, and
\begin{scriptsize}
\begin{align}
 \chi&=\frac{(b_x-i b_y)^2}{32}\frac{\Omega}{\Omega'}(1+\frac{\omega-p}{\Omega'})\left[\Gamma_x^{*}(\omega)-\Gamma_x(-\omega)+\Gamma_x(\Omega'-\omega)-\Gamma_x^{*}(\omega-\Omega') \right]\\
 \beta_1&=\frac{(b_x-i b_y)^2}{16}\frac{\Omega}{\Omega'}(1+\frac{\omega-p}{\Omega'})\left[\Gamma_x(\Omega'-\omega)+\Gamma_x^{*}(\omega-\Omega') \right]\\
 \beta_2&=\frac{(b_x-i b_y)^2}{16}\frac{\Omega}{\Omega'}(1+\frac{\omega-p}{\Omega'})\left[\Gamma_x^{*}(\omega)+\Gamma_x(-\omega) \right]
 \end{align}
\end{scriptsize}
for the XXZ and XXX cases.

The matrix elements of the Fourier coefficients $\Omega'$ in the XYZ are 
\begin{scriptsize}
\begin{align}
 \chi&=\frac{b_x^2}{32}\frac{\Omega}{\Omega'}
 \left\{(1+\frac{\omega-p}{\Omega'})\left[-\Gamma_x(\omega)+\Gamma_x^{*}(-\omega)+\Gamma_x(\Omega'-\omega)-\Gamma_x^{*}(\omega-\Omega') \right]+\right.\nonumber\\&\hspace{2cm}+\left.
 (1-\frac{\omega-p}{\Omega'})\left[-\Gamma_x^{*}(\omega)+\Gamma_x(-\omega)+\Gamma_x^{*}(-\Omega'-\omega)-\Gamma_x(\Omega'+\omega) \right]
 \right\} +"x\leftrightarrow y" +\nonumber\\
 &\hphantom{=}+\frac{b_z^2}{8}\frac{\Omega}{\Omega'}\frac{\omega-p}{\Omega'}\left[\Gamma_z(0)-\Gamma_z^{*}(0)-\Gamma_z(\Omega')+\Gamma_z^{*}(-\Omega) \right]\\
\beta_1&=\frac{b_x^2}{16}\frac{\Omega}{\Omega'}
\left\{(1+\frac{\omega-p}{\Omega'})\left[\Gamma_x(\Omega'-\omega)+\Gamma_x^{*}(\omega-\Omega')\right]-
(1-\frac{\omega-p}{\Omega'})\left[\Gamma_x(\Omega'+\omega)+\Gamma_x^{*}(-\omega-\Omega')\right]
\right\}+ "x\leftrightarrow y"-\nonumber\\
&\hphantom{=}-\frac{b_z^2}{4}\frac{\Omega}{\Omega'}\frac{\omega-p}{\Omega'}\left[\Gamma_z(\Omega')+\Gamma_z^{*}(-\Omega') \right]\\
\beta_2&=\frac{b_x^2}{16}\frac{\Omega}{\Omega'}
\left\{(1+\frac{\omega-p}{\Omega'})\left[\Gamma_x(\omega)+\Gamma_x^{*}(-\omega)\right]-
(1-\frac{\omega-p}{\Omega'})\left[\Gamma_x(-\omega)+\Gamma_x^{*}(\omega)\right]
\right\}+ "x\leftrightarrow y"-\nonumber\\
&\hphantom{=}-\frac{b_z^2}{4}\frac{\Omega}{\Omega'}\frac{\omega-p}{\Omega'}\left[\Gamma_z(0)+\Gamma_z^{*}(0) \right]\\
\delta&=\frac{b_x^2}{32}\left[\left(\frac{\omega-p}{\Omega'}\right)^2-1\right]\left[\Gamma_x(\Omega'+\omega)+\Gamma_x(\Omega'-\omega)+\Gamma_x^{*}(\omega-\Omega')+\Gamma_x^{*}(-\omega-\Omega')\right]+"x\leftrightarrow y"+\nonumber\\
&\hphantom{=}+\frac{b_z^2}{8} \frac{\Omega^2}{\Omega'^2}\left[\Gamma_z(\Omega')+\Gamma_z^{*}(-\Omega)\right]
 \end{align}
\end{scriptsize}
 and the same for the XXZ/XXX cases are given by the substitution $\Gamma_y=\Gamma_x$, $\Gamma_y=\Gamma_z=\Gamma_x$.

\end{document}